\catcode`\@=11

\newif\if@fewtab\@fewtabtrue

{\count255=\time\divide\count255 by 60
\xdef\hourmin{\number\count255}
\multiply\count255 by-60\advance\count255 by\time
\xdef\hourmin{\hourmin:\ifnum\count255<10 0\fi\the\count255}}
\def\ps@draft{\let\@mkboth\@gobbletwo
    \def\@oddhead{}
    \def\@oddfoot
       {\hbox to 7 cm{$\scriptstyle Draft\ version:\ \draftdate$
       \hfil}\hskip -7cm\hfil\rm\thepage \hfil}
    \def\@evenhead{}\let\@evenfoot\@oddfoot}

\def\ceqno{\global\@fewtabfalse
    \ifcase\@eqcnt \def\@tempa{& & &}\or \def\@tempa{& &}
      \or \def\@tempa{&}
      \or\def\@tempa{}\fi\@tempa
{\rm(\theequation)}}

\def\aeqno#1{\global\@fewtabfalse
    \ifcase\@eqcnt \def\@tempa{& & &}\or \def\@tempa{& &}
      \or \def\@tempa{&}
      \or\def\@tempa{}\fi\@tempa
{\rm(\theequation,#1)}}

\def\label#1{\ifnum\draftcontrol=1
 \global\def\draftnote{$\scriptstyle #1$}\fi
 \@bsphack\if@filesw {\let\thepage\relax
   \def\protect{\noexpand\noexpand\noexpand}%
\xdef\@gtempa{\write\@auxout{\string
      \newlabel{#1}{{\@currentlabel}{\thepage}}}}}\@gtempa
   \if@nobreak \ifvmode\nobreak\fi\fi\fi
  \@esphack}

\def\alabel#1#2{\label{#1}\global\@fewtabfalse
    \ifcase\@eqcnt \def\@tempa{& & &}\or \def\@tempa{& &}
      \or \def\@tempa{&}
      \or\def\@tempa{}\fi\@tempa
{\hbox to 3cm{\phantom{\rm(\theequation,#2)}
\draftnote \hfil}\hskip -3cm {\rm(\theequation,#2)}}}

\def\clabel#1{\label{#1}\global\@fewtabfalse
    \ifcase\@eqcnt \def\@tempa{& & &}\or \def\@tempa{& &}
      \or \def\@tempa{&}
      \or\def\@tempa{}\fi\@tempa
{\hbox to 3cm{\phantom{\rm(\theequation)}
\draftnote \hfil}\hskip -3cm{\rm(\theequation)}}}

\def\eqnarray{\def\draftnote{{}}\global\@fewtabtrue
\stepcounter{equation}\let\@currentlabel=\theequation
\global\@eqnswtrue
\global\@eqcnt\z@\tabskip\@centering\let\\=\@eqncr
$$\halign to \displaywidth\bgroup\@eqnsel\hskip\@centering\@eqcnt\z@
  $\displaystyle\tabskip\z@{##}$&\global\@eqcnt\@ne
  \hskip 1\arraycolsep \hfil${##}$\hfil
  &\global\@eqcnt\tw@ \hskip 1\arraycolsep
$\displaystyle\tabskip\z@{##}$
\hfil  \tabskip\@centering&\global\@eqcnt\thr@@\llap{##}\tabskip\z@
\cr}

\def\endeqnarray{\@@eqncr\egroup
      \global\advance\c@equation\m@ne$$\global\@ignoretrue}

\def\@eqnnum{\hbox to 3cm{\phantom{\rm(\theequation)} \draftnote
                         \hfil}\hskip -3cm {\rm(\theequation)}}

\def\@@eqncr{\let\@tempa\relax
    \ifcase\@eqcnt \def\@tempa{& & &}\or \def\@tempa{& &}
      \or \def\@tempa{&}
      \or\def\@tempa{}
\fi\@tempa
\if@eqnsw
\if@fewtab\@eqnnum\fi
\stepcounter{equation}\fi\global
\@eqnswtrue\global\@eqcnt\z@\global\@fewtabtrue\cr}

\def\draftcite#1{\ifnum\draftcontrol=1#1\else{}\fi}

\def\@lbibitem[#1]#2{\item{}\hskip -3cm \hbox to 2cm
{\hfil$\scriptstyle\draftcite{#2}$}\hskip
1cm[\@biblabel{#1}]\if@filesw
     {\def\protect##1{\string ##1\space}\immediate
      \write\@auxout{\string\bibcite{#2}{#1}}}\fi\ignorespaces}

\def\@bibitem#1{\item\hskip -3cm \hbox to 2cm
{\hfil $\scriptstyle\draftcite{#1}$}\hskip 1cm
\if@filesw \immediate\write\@auxout
       {\string\bibcite{#1}{\the\value{\@listctr}}}\fi\ignorespaces}

\def\nsection#1{\section{#1}\setcounter{equation}{0}}

\def\nappendix#1{\vskip 1cm\no{\bf Appendix #1}\def\thesection{#1}
\setcounter{equation}{0}}

\font\tendl=msbm10  scaled \magstep1
\font\sevendl=msbm7 scaled \magstep1
\font\fivedl=msbm5 scaled \magstep1
\font\tengl=eufm10  scaled \magstep1
\font\sevengl=eufm7 scaled \magstep1
\font\fivegl=eufm5 scaled \magstep1

\newfam\dlfam  
\textfont\dlfam=\tendl \scriptfont\dlfam=\sevendl
\scriptscriptfont\dlfam=\fivedl
\newfam\glfam  
\textfont\glfam=\tengl \scriptfont\glfam=\sevengl
\scriptscriptfont\glfam=\fivegl

\def\draftdate{\number\month/\number\day/\number\year\ \ \ \hourmin }

\global\def\draftcontrol{0}
\catcode`\@=12
\def\tilde{\widetilde}
\def\hat{\widehat}
\documentclass[aps,amsfonts,amsmath,amssymb,floatfix,nofootinbib]{revtex4}
\usepackage{epsfig}
\usepackage{rotating}
\usepackage[centerlast]{subfigure}
\usepackage{bm}
\usepackage{amssymb}
\renewcommand{\theequation}{\thesection.\arabic{equation}}
\newcommand{\ii}{\mathrm{i}}
\newcommand{\be}{\begin{eqnarray}}
\newcommand{\en}{\end{eqnarray}\vs 0.5 cm}

\newcommand{\Id}{{I\hspace{-0.04cm}d}}

\newcommand{\no}{\noindent}
\newcommand{\vs}{\vskip}

\newcommand{\qq}{\begin{eqnarray}}

\newcommand{\ee}{{\rm e}}

\newcommand{\qqq}{\end{eqnarray}}

\newcommand{\tr}{\hbox{tr}}
\newcommand{\pf}{{\rm pf}}

\newcommand{\hol}{{h\hspace{-0.01cm}ol}}
\newcommand{\Hol}{{H\hspace{-0.02cm}ol}}

\newcommand{\CB}{{\cal B}}

\newcommand{\CD}{{\cal D}}
\newcommand{\CE}{{\cal E}}

\newcommand{\CG}{{\cal G}}

\newcommand{\CL}{{\cal L}}

\newcommand{\WZ}{{W\hspace{-0.06cm}Z}}
\newcommand{\SC}{{C\hspace{-0.05cm}S}}

\newcommand{\KM}{{K\hspace{-0.06cm}M}}
\begin{document}
\title{$2d\,$ Fu-Kane-Mele invariant as Wess-Zumino action\\ 
of the sewing matrix}
\author{Krzysztof Gaw\c{e}dzki\footnote{directeur de recherche \'em\'erite, 
\,email: kgawedzk@ens-lyon.fr}}
\affiliation{Universit\'e de Lyon, ENS de Lyon, Universit\'e Claude Bernard, 
CNRS\\ Laboratoire de Physique, F-69342 Lyon, France\\
}

\date{\today}

\maketitle

\centerline{\bf Abstract}
\vskip 0.4cm

\quad\ \parbox[t]{15cm}{We show that the Fu-Kane-Mele invariant of the $2d$ 
time-reversal-invariant crystalline insulators is equal to the properly 
normalized Wess-Zumino action of the  so-called sewing matrix field 
defined on the Brillouin torus. Applied to $3d$, the result permits 
a direct proof of the known relation between the strong Fu-Kane-Mele 
invariant and the Chern-Simons action of the non-Abelian Berry connection 
on the bundle of valence states.}

\vskip 1.3cm

\nsection{Introduction}
\label{sec:intro}

\noindent $2d$ crystalline insulators invariant under odd time-reversal 
are classified by the Kane-Mele $\,\mathbb Z_2$-valued bulk invariant 
introduced in \cite{KaneMele} and rewritten in \cite{FuKane06} 
in a form that will be used here. In \cite{FKM}, a similar $\mathbb Z_2$-valued 
invariant, called strong, was defined for $3d$ time-reversal invariant 
(TRI) crystalline insulators. The physical importance of such 
invariants relies on the fact that their nontrivial value guaranties 
the existence of robust massless edge modes in finite samples of TRI 
crystals. In \cite{QHZ}, the origin of the $\mathbb Z_2$-valued 
invariants in $3d$ and $2d$ was traced back to the integer-valued 
$2^{\rm nd}$ Chern number of the vector bundle of valence Bloch states 
of $4d$ crystalline insulators without TRI. This was achieved in a chain
of dimensional reductions from 4 space dimensions to 3 and then from 
3 to 2. In particular, a $\,\mathbb Z_2$-valued invariant 
of $3d$ TRI insulators was expressed as the Chern-Simons (CS) 
action functional of the non-Abelian Berry connection of the $3d$ 
valence bundle. The CS action, when normalized to change by even integers 
under gauge transformations, is forced by TRI to take integer 
values giving rise to a $\,\mathbb Z_2$-valued index. An indirect proof 
that such an index coincides with the strong invariant defined in \cite{FKM} 
was given in \cite{FM}, where it was shown that both represent the same 
$\,\mathbb Z_2$ subgroup of the Real K-theory group $\,K\hspace{-0.04cm}
R^{-4}\,$ of the $3d$ Brillouin torus. 
\vskip 0.1cm

A continuation of the line of thought of \cite{QHZ} permits 
to define an invariant of $2d$ TRI crystals using another topological 
action functional: the Wess-Zumino action (WZ) of the so called sewing-matrix 
field defined on the $2d$ Brillouin torus. When properly normalized, 
such an action is defined modulo even integers and it takes integer 
values when calculated on the sewing-matrix field. The main result 
of this note is a direct proof that the $2d$ $\,\mathbb Z_2$-valued invariant 
obtained this way coincides with the one defined in \cite{FuKane06}. 
Sec.\,2 is devoted to the precise statement of the result and Sec.\,3 
to its proof employing the technique of bundle gerbes, particularly suitable 
for the calculation of WZ actions. In Sec.\,3, we show, basing on \cite{QHZ}, 
that our $2d$ result also permits to directly prove the equality 
between the CS action of the $3d$ Berry connection and the strong 
$\,\mathbb Z_2$ invariant of \cite{FKM}.

\nsection{Statements of the main result} 
\label{sec:mainresult}

\noindent Consider a smooth family of Hermitian $N\times N$ matrices 
$\,H(k)\,$ parameterized by $\,k\in\mathbb R^d\,$ and satisfying the relations 
\qq
&&H(k+b)\hspace{0.03cm}\ =H(k)\qquad\hspace{0.02cm}{\rm for}
\quad b\in2\pi\mathbb Z^d\,,\label{1}\\
&&\theta H(k)\theta^{-1}=H(-k)\quad\ {\rm for}\quad{\rm an\ antiunitary} 
\ \,\theta:\mathbb C^N\rightarrow\mathbb C^N\quad{\rm with}\quad \theta^2=-I\,.\label{2}
\qqq
Such families appear in the context of two-dimensional lattice tight-binding 
TRI systems where $\,H(k)\,$ describe the Bloch Hamiltonians. Condition (\ref{1}) 
means that such Hamiltonians are effectively defined on the $d$-dimensional 
Brillouin torus $\,\mathbb T^d=\mathbb R^d/(2\pi\mathbb Z^d)$. The map $\,\theta\,$ 
realizes the odd time reversal and (\ref{2}) expresses the time-reversal symmetry 
of the system. The existence of $\,\theta\,$ squaring to $\,-I\,$ requires that 
$\,N\,$ be even.
\vskip 0.1cm

Suppose that there exists $\,\epsilon_F\in\mathbb R\,$ that is not in the spectrum 
of $\,H(k)\,$ for all $\,k$. Such a situation corresponds to systems that 
are insulators in the fermionic second-quantized ground state that fills all 
1-particle eigenstates of $\,H(k)\,$ with eigenvalues $\,<\epsilon_F$, \,called
the valence-band states. \,Let $\,P(k)\,$ be the spectral projectors of $\,H(k)\,$ 
corresponding to such eigenvalues. The projectors $\,P(k)\,$ depend smoothly 
on $\,k\ {\rm mod}\ 2\pi\mathbb Z\,$ and satisfy the relation
\qq
\theta P(k)\theta^{-1}=P(-k)\label{4}
\qqq
following from (\ref{2}). The ranges of $\,P(k)\,$ form a 
vector sub-bundle $\,\CE\,$ of the trivial bundle $\,\mathbb T^d\times\mathbb C^N\,$ that
will be called the valence subbundle. We shall denote by $n$ its rank, i.e. the dimension of 
the ranges of projectors $\,P(k)$. \,Necessarily, $\,n\,$ is even due to the time-reversal 
symmetry (\ref{4}). In $\,d=2,3$, \,the vector bundle $\,\CE\,$ is trivializable. 
This follows from the vanishing of its first Chern number(s), \,another consequence 
of (\ref{4}) \cite{Panati}. \,The trivializability of $\,\CE\,$ means that there exists 
a smooth family $\,(e_i(k))_{i=1}^n\,$ of vectors in $\,\mathbb C^N$, such that 
$\,e_i(k)=e_i(k+b)\,$ for $\,b\in2\pi\mathbb Z^d$, and, for each $\,k$, $\,(e_i(k))\,$ form 
an orthonormal basis of the range of $\,P(k)$. In what follows, a prominent role will 
be played by the $n\times n$ unitary ``sewing matrices'' $\,w(k)\,$ with the entries
\qq
w_{ij}(k)=\big\langle e_i(-k)|\theta e_j(k)\big\rangle\,,
\label{6}
\qqq
depending smoothly on $\,k\ {\rm mod}\ 2\pi\mathbb Z\,$ and obeying the relation
\qq
w(-k)=-w(k)^T\,.\label{TRw}
\qqq
That relation implies that the matrix $\,w(k)\,$ is antisymmetric
at points of $\,\mathbb T^d\,$ where $\,k=-k\ {\rm mod}\ 2\pi\mathbb Z^d$, 
\,the so called TRIM (time-reversal invariant (quasi-)momenta). There are $\,2^d\,$
such points in $\,\mathbb T^d$. \,It also follows from (\ref{TRw}) that 
$\,\det{w(k)}=\det{w(-k)}\,$ for all $\,k$. \,The latter relation implies 
that $\,\det{w(k)}\,$ does not wind along the basic cycles of $\,\mathbb T^d\,$
so that one may define a smooth function $\,\ln{\det{w(k)}}=\ln{\det{w(-k)}}\,$ on 
$\,\mathbb T^d\,$ uniquely up to a global additive constant in $\,2\pi\ii\mathbb Z$.
In particular, one may define smooth roots $\,\sqrt[p]{\det{w(k)}}=\exp[\frac{1}{p}
\ln{\det{w(k)}}]\,$ over $\,\mathbb T^d\,$ up to a global factor equal to
a $p^{\rm th}$ root of unity.
\vskip 0.1cm

In \cite{KaneMele}, Kane and Mele realized that in dimension $\,d=2$, \,there is 
an obstruction $\,\KM\in\mathbb Z_2\,$ whose non-zero value forbids that 
$\,(e_i(k))\,$ be composed of Kramers' pairs satisfying the conditions
\qq
e_{2i}(-k)=\theta e_{2i-1}(k)
\label{5}
\qqq
for all $\,k$. \,Note that relations (\ref{5}) demands that 
$\,w(k)\,$ be composed of $\,k$-independent $2\times 2$ matrices 
\qq
\Big(\begin{matrix}0&{-1}\cr1&0\end{matrix}\Big)
\qqq 
placed diagonally. In \cite{FuKane06}, the Kane-Mele obstruction $\,\KM\,$ 
to achieve such a form of $\,w(k)\,$ was expressed with the help of arbitrary
family of sewing matrices via the multiplicative relation
\qq
(-1)^{\KM}=\prod\limits_{{\rm TRIM}\,\in\,\mathbb T^2}
\frac{\sqrt{\det{w(k)}}}{\pf\,{w(k)}}
\label{FKM2}
\qqq
where the product is over the four TRIM in $\,\mathbb T^2$, $\,\,\pf\,$ denotes 
the pfaffian defined for antisymmetric matrices and $\,\sqrt{\det{w(k)}}\,$ is
defined as above. Since $\,\sqrt{\det{w(k)}}/\pf\,{w(k)}\,$ squares to $\,1$, 
\,the right hand side of (\ref{FKM2}) is $\,\pm1$. \,It is independent
of the global sign ambiguity in the definition of $\,\sqrt{\det{w(k)}}\,$ and may 
be shown \cite{FM} to be independent of the choice of the trivialization $\,(e_i(k))$, 
\,determining uniquely $\,\KM\in\mathbb Z_2$. It was rigorously shown in 
\cite{DNG} and in \cite{FMP} that $\,\KM\,$ is the only obstruction
for trivializing the $2d$ TRI valence bundle with Kramers' pairs.
\vskip 0.1cm

The first part of the present note is devoted to the proof of the following 
result announced in \cite{Geneva}
\vskip 0.4cm

\noindent{\bf Theorem.}
\vskip -1cm
\qq
(-1)^\KM\,=\,\exp[\ii S_\WZ(w)]
\label{topr}
\qqq
\vskip 0.2cm

\noindent that establishes the equality between the Kane-Mele index 
$\,\KM\in\mathbb Z_2\,$ and the two-dimensional Wess-Zumino (WZ) action
$\,S_\WZ(w)\,$ divided by $\pi$ of the unitary-group-valued field 
$\,\mathbb T^2\ni k\mapsto w(k)\,\in U(n)$. 
\,The WZ action in question is defined following Witten's prescription 
\cite{Wit}: \,one extends the field $\,w\,$ to a $\,U(n)$-valued map $\,W\,$
on an oriented 3-dimensional manifold $\,\CB\,$ with boundary 
$\,\partial\CB =\mathbb T^2$, \,demanding that $\,W|_{\partial\CB}=w$, \,and 
one sets
\qq
S_\WZ(w)\,=\,\int_\CB W^*H,
\label{WZ}
\qqq
where $\,H\,$ is a closed bi-invariant 3-form on the unitary group $\,U(n)$,
\qq
H\,=\,\frac{_1}{^{12\pi}}\,\tr\,(g^{-1}dg)^3
\label{H}
\qqq
normalized so that its $3$-periods are in $\,2\pi\mathbb Z$.
An extension $\,W\,$ of $\,w\,$ always exists for a suitable $\,\CB\,$ 
and the right hand side of (\ref{WZ}) is well defined modulo $\,2\pi$.
\,This makes $\,\frac{1}{\pi}S_\WZ(w)\,$ defined modulo $\,2$ and the 
WZ Feynman amplitude $\,\exp[\ii S_\WZ(w)]\,$ uniquely defined.
\vskip 0.4cm

\noindent{\bf Remark.} \ It is not difficult to proof using
the basic properties of the WZ action, see Appendix, that the right 
hand side of (\ref{topr}) is equal to $\,\pm1$, \,does not depend on 
the choice of the trivialization $\,(e_i(k))\,$ of the valence bundle 
$\,\CE\,$ and is invariant under smooth deformations of $\,w\,$ 
preserving the symmetry (\ref{TRw}), so that the formula (\ref{topr}) 
renders more transparent the topological nature of the $2d$ Fu-Kane-Mele 
invariant.
\vskip 0.4cm

It will be more convenient in the sequel to remove a $U(1)$ contribution
from $\,w\,$ and to work with an $\,SU(n)$-valued field
\qq
\tilde w(k)=(\sqrt[n]{\det{w(k)}})^{-1}w(k)
\qqq
on $\,\mathbb T^2\,$ using one of the smooth $n^{\rm th}$ roots defined above.  
\,One may choose an extension $\,W\,$ of $\,w\,$ such that $\,W=D\widetilde W$, 
\,where $\,D:S\mapsto U(1)\,$ extends $\,(\sqrt[n]{\det{w(k)}})^{-1}\,$ and 
$\,\widetilde W:S\mapsto SU(m)\,$ extends $\,\widetilde w$. \,By the 
formula (\ref{prfmla}) in Appendix,
\qq
W^*H=(D\widetilde W)^*H=D^*H+\widetilde W^*H+3d[(D^{-1}dD)\,\tr\,
(\widetilde Wd\widetilde W^{-1})]
=\widetilde W^*H   
\qqq
because $\,D^*H=0\,$ for dimensional reasons and 
$\,\tr(\widetilde Wd\widetilde W^{-1})=0\,$ because the 1-form 
$\,\widetilde Wd\widetilde W^{-1}\,$ takes values in traceless matrices. 
It follows that
\qq
\exp[\ii S_\WZ(w)]=\exp[\ii S_\WZ(\widetilde w)]\,.
\qqq 
Note that $\,\widetilde w(k)\,$ still satisfies the relation
(\ref{TRw}), \,i.e. 
\qq
\widetilde w(-k)=-\widetilde w(k)^T\label{TRwt}
\qqq
and that at the TRIM, 
\qq
\pf\,\widetilde w(k)\,=\,\frac{\pf\,w(k)}{\sqrt{\det{w(k)}}}\,=\,\frac{1}{\pf\,
\widetilde w(k)} 
\qqq
so that
\qq
(-1)^\KM=\prod\limits_{{\rm TRIM}\,\in\,\mathbb T^2}\pf\,\widetilde w(k)\,.
\qqq
Hence the Theorem above may be reduced to the following result that localizes 
the WZ amplitude of $\,\tilde w\,$ at the TRIM: 
\vskip 0.4cm

\noindent{\bf Proposition 1.}
\vskip -1.05cm
\qq
\exp[\ii S_\WZ(\widetilde w)]\ =\hspace{-0.1cm}\prod\limits_{{\rm TRIM}\,\in\,\mathbb T^2}
\pf\,\widetilde w(k)\,.
\label{topr2}
\qqq

\nsection{Wess-Zumino amplitude as a gerbe holonomy} 
\label{sec:gerbeholonomy}

\noindent In order to prove Proposition 1, we shall reinterpret 
the WZ amplitude $\,\exp[\ii S_\WZ(\widetilde w)]\,$ as the holonomy of a bundle 
gerbe\footnote{All bundle gerbes and 
line bundles considered below come equipped with a Hermitian structure 
and a Hermitian connection and their isomorphisms are be assumed to 
respect those structures.} $\,\CG\,$ over the group $\,SU(n)\,$ 
\cite{Murray,Chatt,GR}. \,That will provide a local expression for the WZ 
amplitude of $\,\widetilde w\,$ with multiple cancellations, allowing 
at the end its localization at the TRIM. Loosely speaking, 
(bundle) gerbes are structures one degree higher than line bundles. Their 
holonomies are defined along closed surfaces rather than along loops and 
their curvatures are closed 3-forms rather than closed 2-forms. The gerbe 
$\,\CG\,$ over $\,SU(n)$, \,called basic, is characterized, up to isomorphism, 
by its curvature equal to the 3-form $\,H\,$ of (\ref{H}) (restricted 
to the special unitary group). We shall employ a construction of $\,\CG\,$ 
from \cite{GR}, that we briefly recall here, giving subsequently a local 
expression for the holonomy of $\,\CG$.
\vskip 0.1cm

Let $\,\lambda_i,$\ $i=1,\dots,n-1$, \,be the standard choice for simple 
weights of the Lie algebra $\,su(n)\,$ given by the diagonal $n\times n$ 
matrices with the entries 
\qq
\big(\mathop{\frac{_{n-i}}{^n},\dots,\frac{_{n-i}}{^n}}\limits_{
i\ {\rm times}},\frac{_{-i}}{^n},\dots,\frac{_{-i}}{^n}\big)\,
\qqq 
and let $\,\lambda_0=0$. \,Below, $\,\lambda_{ij}\,$ will stand for 
the difference $\,\lambda_j-\lambda_i\,$ for $\,i,j=0,\dots,n-1$. 
\,One chooses a covering $\,(O_i),$\ $i=0,\dots,n-1$, \,of $\,SU(n)\,$ 
composed of open subsets
\qq
O_i\,=\,\Big\{g=\gamma\,\ee^{2\pi\ii\tau}\gamma^{-1}\ \big|\ 
\gamma\in SU(n),\ \tau=
\sum\limits_{j=0}^{n-1}\tau_j\lambda_j\ \,{\rm with}\,\,\,\,
0\leq\tau_j,\ \sum_j\tau_j=1,\ \tau_i>0\Big\}
\qqq
equipped with smooth 2-forms 
\qq
B_i(g)\,
=\,\frac{_1}{^{4\pi}}\,\tr((\gamma^{-1}d\gamma)\,\ee^{2\pi\ii\tau}
(\gamma^{-1}d\gamma)\,\ee^{-2\pi\ii\tau})\,+\,\ii\,\tr((\tau-\lambda_i)
(\gamma^{-1}d\gamma)^2
\qqq
such that $\,dB_i=H|_{O_i}$. \,Let $\,O_i\cap O_j\equiv O_{ij}\,$ be a double
intersection of sets of the covering. Then 
\qq
B_{ij}(g)=B_j(g)-B_i(g)=-\,\tr\,\lambda_{ij}(\gamma^{-1}d\gamma)^2
\qqq
is a closed 2-form over $\,O_{ij}$. \,If 
\qq
g=\gamma\,\ee^{2\pi\ii\tau}\gamma^{-1}=\gamma\gamma_0^{-1}\,
\ee^{2\pi\ii\tau}\gamma_0\gamma^{-1}\in O_{ij}
\qqq
then, necessarily, $\,\gamma_0\in G_{ij}$, \,where
\qq
G_{ij}=\{\gamma_0\in SU(n)\,|\,\gamma_0\lambda_{ij}\gamma_0^{-1}=\lambda_{ij}\}.
\qqq
All groups $\,G_{ij}\,$ are connected and they contain the Cartan subgroup 
$\,T\subset SU(n)\,$ composed of the diagonal $\,SU(n)\,$ matrices.
Let $\,\chi_{ij}:G_{ij}\rightarrow U(1)\,$ be the character of $\,G_{ij}\,$ 
defined by the relations
\qq
\chi_{ij}(I)=1\,,\qquad
d\ln\chi_{ij}(\gamma_0)=\tr(\lambda_{ij}\gamma_0^{-1}d\gamma_0)\,.
\label{chiij}
\qqq
In particular, for a real traceless diagonal matrix $\,\phi$,
\qq
\chi_{ij}(\ee^{\ii\phi})\,=\,\ee^{\ii\,\tr(\lambda_{ij}\phi)}\,.
\label{tdm}
\qqq
Over the double intersections $\,O_{ij}\,$ one considers
the line bundles $\,L_{ij}\,$ composed of the equivalence classes
$\,[\gamma,\zeta]_{ij}\,$ with $\,\zeta\in\mathbb C\,$ such that
\qq
(\gamma,\zeta)\ \sim_{ij}\,(\gamma\gamma_0^{-1},\chi_{ij}(\gamma_0)\zeta)
\label{eqrel}
\qqq
for $\,\gamma_0\in G_{ij}$. The line bundle $\,L_{ij}\,$ comes equipped
with the Hermitian structure $|[\gamma,\zeta]_{ij}|=|\zeta|\,$ and
the Hermitian connection induced by the connection form 
\qq
A_{ij}(g)=\tr(\lambda_{ij}\gamma^{-1}d\gamma)
\qqq 
whose curvature is given by the 2-form $\,B_{ij}$. 
\,Finally, over the triple intersections $\,O_{ijk}\,$ there exist
line-bundle isomorphisms $\,L_{ij}\otimes L_{jk}\mathop{\longrightarrow}\limits^{t_{ijk}}L_{ik}\,$
defined by
\qq
t_{ijk}\big([\gamma,\zeta]_{ij}\otimes[\gamma,\zeta']_{jk}\big)=
[\gamma,\zeta\zeta']_{ik}
\qqq
that behave in an associative way over the quadruple intersections 
$\,O_{ijkl}\,$ so that
\qq
t_{ikl}\circ(t_{ijk}\otimes\Id_{L_{kl}})=t_{ijl}\circ(\Id_{L_{ij}}\otimes t_{jkl})
\,.
\label{assoc}
\qqq
The isomorphisms $\,t_{iij}\,$ and $\,t_{iji}\,$ permit to canonically identify 
$\,L_{ii}\,$ with the trivial bundle $\,O_i\times\mathbb C\,$ and 
$\,L_{ji}\,$ with the line bundle $\,L_{ij}^{-1}\,$ dual to $\,L_{ij}$. 
\,The basic gerbe $\,\CG\,$ over $\,SU(n)\,$ is defined 
by the structure described above, \,i.e. $\,\CG=((O_i),(B_i),(L_{ij}),
(t_{ijk}))$. 
\vskip 0.1cm

Bundle gerbes over a manifold $\,M\,$ allow to define a $\,U(1)$-valued 
holonomy of smooth maps $\,\phi:\Sigma\rightarrow M\,$ from a closed oriented 
surface $\,\Sigma\,$ to $\,M$. In particular, for the basic gerbe $\,\CG\,$ 
over $\,SU(n)\,$
the holonomy $\,\Hol_\CG(\phi)\,$ is identified with the WZ-amplitude 
$\,\exp[\ii S_{\WZ}(\phi)]$, \,and the use of the gerbe structure allows to
write a local expression for the latter as described in \cite{GR}, see also 
the earlier works \cite{Alvarez,KG}. This is done in the following way.
One chooses a triangulation  $\,\{(c),(b),(v)\}\,$ of $\,\Sigma$, \,composed of 
triangles $\,c$, \,edges $\,b\,$ and vertices $\,v$, \,that is sufficiently fine 
so that it is possible to fix indices $\,i_c,i_b,i_v\,$ satisfying
\qq
\phi(c)\subset O_{i_c},\quad\phi(b)\subset O_{i_b},\quad\phi(v)\in O_{i_v}.
\qqq
For each $\,b\subset c$, \,let us denote by $\,\hol_{L_{i_ci_b}}(\phi|_b)\,$
the parallel transport in the line bundle $\,L_{i_ci_b}\,$ along the
(open) curve $\,\phi|_b\,$ for $\,b\,$ oriented as the boundary edge of the triangle 
$\,c$. \,Thus 
\qq
\hol_{L_{i_ci_b}}(\phi|_b)\in\mathop{\otimes}\limits_{v\in b}
(L_{i_ci_b})_{\phi(v)}^{\pm 1}\,,
\label{hol}
\qqq
where $\,(L_{ij})_g^{\pm1}\,$ denotes the fiber of $\,L_{ij}\,$ over
$\,g\in O_{ij}\,$ or its dual, and on the right hand side the plus (the minus) 
sign is chosen if the vertex $\,v\,$ is the end point (the starting point)
of the oriented edge $\,b$. \,The local expression for the gerbe holonomy
takes the form  
\qq
\Hol_\CG(\phi)\,=\,\exp\Big[\ii\sum\limits_c\int_c\phi^*B_{i_c}\Big]
\mathop{\otimes}\limits_{\substack{(b,c)\\b\subset c}}\hol_{L_{i_ci_b}}(\phi|_b)\,.
\label{Hol}
\qqq
As it stands, the right hand side, is an element of the line
\qq
\mathop{\otimes}\limits_{\substack{(b,c)\\b\subset c}}\Big(
\mathop{\otimes}\limits_{v\in b}
(L_{i_ci_b})_{\phi(v)}^{\pm 1}\Big)\,=\,
\mathop{\otimes}\limits_v\Big(\mathop{\otimes}\limits_{\substack{(b,c)\\
v\in b\subset c}}(L_{i_ci_b})_{\phi(v)}^{\pm 1}\Big)
\qqq
that may be canonically identified with $\,\mathbb C\,$ using the
isomorphisms $\,t_{ijk}$, \,see \cite{GR} or \cite{Geneva} for more
details. Such an identification defines $\,\Hol_\CG(\phi)\,$ as a 
(modulus 1) complex number that is independent of the choice of 
the triangulation of $\,\Sigma\,$ and the assignments $\,(i_c,i_b,i_v)$,
\,as may be easily checked.     
\vskip 0.1cm

Above we assumed that $\,\partial\Sigma=\emptyset$. \,In the case when 
$\,\partial\Sigma\not=\emptyset\,$ and is composed of a closed loop, 
the right hand side of (\ref{Hol}) cannot be canonically viewed as an element 
of $\,\mathbb C\,$ but only as an element of the line
\qq
\mathop{\otimes}\limits_{\substack{(v,b)\\v\in b\subset\partial\Sigma}}
(L_{i_vi_b})^{\pm1}_{\phi(v)}\,\equiv\,(\CL^\CG)_{\phi|_{\partial\Sigma}}
\label{defline}
\qqq
with the same sign convention that in (\ref{hol}). Such lines may be 
canonically identified for different triangulations
of $\,\partial\Sigma\,$ (composed of edges $\,b\,$ and vertices $\,v$)
\,defining the fibers $\,(\CL^\CG)_{\phi|_{\partial\Sigma}}\,$
of the transgression line bundle $\,\CL^\CG\,$ over the loop group $\,LSU(n)\,$
canonically associated to the gerbe $\,\CG\,$ \cite{GR}. Hence
\qq
\Hol_\CG(\phi)\,\in\,(\CL^\CG)_{\phi|_{\partial\Sigma}}
\qqq
in this case. The lines $\,(\CL^\CG)_\varphi\,$ related to loops $\,\varphi\,$
differing by an orientation-preserving reparameterization are canonically
isomorphic and those related by an orientation-reversing reparameterization
are canonically dual. More generally, if $\,\partial\Sigma=\sqcup S_a\,$ 
is composed of loops $\,S_a\,$ then
\qq
\Hol_\CG(\phi)\,\in\,\mathop{\otimes}\limits_{a}
(\CL^\CG)_{\phi|_{S_a}}\equiv\,(\CL^\CG)_{\phi|_{\partial\Sigma}}\,,
\label{rhtm}
\qqq  
where the rightmost term is the shorthand notation.  
\vskip 0.1cm

Now suppose that $\,n=2m\,$ is even and consider the map 
$\,g\,\mathop{\longmapsto}\limits^r\,-g^T\,$ on $\,SU(n)$. 
If $\,g=\gamma\,\ee^{2\pi\ii\tau}\gamma^{-1}\in O_i\,$ then
\qq
r(g)=(\gamma^{-1})^T{\omega}^{-1}\ee^{2\pi\ii\tau^r}{\omega}\,\gamma^T
\qquad
{\rm for}\qquad
{\omega}=\bigg(\begin{matrix}0&I_m\cr
-I_m&0\end{matrix}\bigg),
\label{r(g)}
\qqq
where $\,I_m\,$ stands for the unit $m\times m$ matrix,
and $\,\ee^{2\pi\ii\tau^r}=-\omega\,\ee^{2\pi\ii\tau}\omega^{-1}$ for
$\,\tau^r=\sum\tau_j\lambda_{j^r}$, \,where $\,0\leq j^r\leq n-1$, 
$\,j^r=j+m\ {\rm mod}\ n$. \,The expression for $\,\tau^r\,$
results from the relation
\qq
{\omega}\lambda_j\,{\omega}^{-1}=\lambda_{j^r}-\lambda_m
\label{ccos}
\qqq
that is straightforward to check. It follows that if $\,g\in O_i\,$ then 
$\,r(g)\in O_{i^r}$, \,i.e. $\,r\,$ maps $\,O_i\,$ into $\,O_{i^r}$.  
\,Since the map $\,j\mapsto j^r\,$ is an involution on the set 
$\,\{0,1,\dots,n-1\}$, $\,r(O_i)=O_{i^r}$.
\,A simple calculation using the invariance of trace 
under the transposition shows that
\qq
r^*B_{i^r}&=&
\frac{_1}{^{4\pi}}\,\tr(({\omega}\gamma^Td((\gamma^{-1})^T{\omega}^{-1}))
\,\ee^{2\pi\ii\hat\tau}({\omega}\gamma^Td((\gamma^{-1})^T{\omega}^{-1}))\,
\ee^{-2\pi\ii\hat\tau})\cr
&&+\,\ii\,\tr((\tau-\lambda_{i^r})
({\omega}\gamma^Td((\gamma^{-1})^T{\omega}^{-1}))^2\cr\cr
&=&\frac{_1}{^{4\pi}}\,\tr((\gamma^Td((\gamma^{-1})^T))
\,\ee^{2\pi\ii\tau}(\gamma^Td((\gamma^{-1})^T))\,\ee^{-2\pi\ii\tau})
+\,\ii\,\tr((\tau-\lambda_i)
(\gamma^Td((\gamma^{-1})^T))^2\cr\cr
&=&-\frac{_1}{^{4\pi}}\,\tr(\ee^{-2\pi\ii\tau}((d\gamma^{-1})\gamma)
\,\ee^{2\pi\ii\tau}((d\gamma^{-1})\gamma)\,
-\,\ii\,\tr(((d\gamma^{-1})\gamma)^2(\tau-\lambda_i)\cr\cr
&=&-\frac{_1}{^{4\pi}}\,\tr((\gamma^{-1}d\gamma)
\,\ee^{2\pi\ii\tau}(\gamma^{-1}d\gamma)\,\ee^{-2\pi\ii\tau})\,
-\,\ii\,\tr((\tau-\lambda_i)(\gamma^{-1}d\gamma)^2\,=\,
-B_i\,.
\label{BrB}
\qqq
\vskip 0.2cm

\noindent{\bf Lemma 1.} \ There are line-bundle isomorphisms 
$\,\nu:L_{ji}\rightarrow r^*L_{i^rj^r}\,$ 
defined by
\qq
\nu([\gamma,\zeta]_{ji})\,
=\,[(\gamma^{-1})^T\omega^{-1},\zeta]_{i^rj^r}\,.
\label{nu}
\qqq
that intertwine the groupoid multiplication.
\vskip 0.2cm

\noindent{\bf Proof of Lemma 1.} \ First we have to check that the 
definition of $\,\nu\,$ is independent of the choice of 
representatives of the equivalence classes. Indeed, for
$\,(\gamma,\zeta)\sim_{ji}(\gamma\gamma_0^{-1},\chi_{ji}(\gamma_0)\zeta)$,
\qq
((\gamma\gamma_0^{-1})^{-1})^T\omega^{-1}=(\gamma^{-1})^T\gamma_0^T\omega^{-1}
=(\gamma^{-1})^T\omega^{-1}\omega\gamma_0^T\omega^{-1}
\qqq
and
\qq
(\omega\gamma_0^T\omega^{-1})^{-1}\lambda_{i^rj^r}
\omega\gamma_0^T\omega^{-1}=\omega(\gamma_0^T)^{-1}\lambda_{ij}\gamma_0^T\omega^{-1}
=\omega(\gamma_0\lambda_{ij}\gamma_0^{-1})^T\omega^{-1}=\omega\lambda_{ij}
\omega^{-1}=\lambda_{i^rj^r}&
\qqq
so that $\,(\omega\gamma_0^T\omega^{-1})^{-1}\in G_{i^rj^r}$. \,Using (\ref{ccos})
and the defining properties (\ref{chiij}) of the characters $\,\chi_{ij}$, 
\,one easily verifies that
\qq
\chi_{i^rj^r}((\omega\gamma_0^T\omega^{-1})^{-1})=
\chi_{ji}(\gamma_0)\,.
\qqq
Hence
\qq
((\gamma^{-1})^T\omega^{-1},\zeta)\,\sim_{i^rj^r}\,
((\gamma^{-1})^T\omega^{-1}(\omega\gamma_0^T\omega^{-1}),
\chi_{i^rj^r}((\omega\gamma_0^T\omega^{-1})^{-1})\zeta)=
((\gamma\gamma_0^{-1})^{-1})^T\omega^{-1},\chi_{ji}(\gamma_0)\zeta)\qquad
\qqq
implying that $\,\nu\,$ is well defined. The identity
\qq
\tr(\lambda_{i^rj^r}\omega\gamma^Td(\gamma^{-1})^T\omega^{-1})=
\tr(\lambda_{ij}\gamma^Td(\gamma^{-1})^T)=\tr(\lambda_{ij}(d\gamma^{-1})\gamma)=
\tr(\lambda_{ji}\gamma^{-1}d\gamma)
\qqq
shows that $\,\nu\,$ intertwines the connections. Clearly, it also 
intertwines the Hermitian structures and the groupoid multiplication. 
\vskip 0.3cm

It follows from Lemma 1 that the isomorphisms $\,\nu\,$ 
intertwine also the parallel transport:
\qq
\hol_{L_{i^rj^r}}(r\circ\phi|_b)=
\hol_{r^*L_{i^rj^r}}(\phi|_b)=
\nu\Big(\hol_{L_{ji}}(\phi|_b)\Big),
\label{holnu}
\qqq
where on the right hand side $\,\nu\,$ is understood to act on
each tensor factor of an element of 
$\,\mathop{\otimes}\limits_{v\in b}(L_{ji})_{\phi(v)}^{\pm1}$.

\nsection{Proof of Proposition 1}
\label{sec:proofProp1}

\noindent Equipped with the gerbe technology, we shall calculate the holonomy
of the basic gerbe $\,\CG\,$ over $\,SU(n)\,$ in the case when 
$\,\Sigma=\mathbb T^2\,$ and $\,\phi=\tilde w\,$ satisfies
relation (\ref{TRwt}) that may be rewritten as the identity
\qq
r\circ\widetilde w=\widetilde w\circ\vartheta
\label{rwt}
\qqq
for the involution $\,\vartheta\,$ of $\,\mathbb T^2\,$ induced
by the map $\,k\mapsto-k$. \,Let $\,\mathbb T^2_+\,$
be the closure of the right half of $\,\mathbb T^2$, \,i.e. 
its  part corresponding to points $\,k=(k_1,k_2)\,$ with $\,0\leq k_1\leq\pi$,
\,and $\,\mathbb T^2_-=\vartheta(\mathbb T^2_+)\,$ be the closure of the
complementary part of $\,\mathbb T^2$, \,see Fig.\,1. Let us choose a 
(sufficiently fine) triangulation of $\,\mathbb T^2\,$ that is symmetric 
under $\,\vartheta\,$ and that restricts to triangulations of 
$\,\mathbb T^2_\pm\,$ and contains the TRIM as vertices. For such a 
triangulation, we shall choose a maximally symmetric assignment of indices 
satisfying 
\qq
i_{\vartheta(c)}=i^r_c\,,\quad\ i_{\vartheta(b)}=i^r_b\,,\quad\ 
i_{\vartheta(v)}=i^r_v
\label{iii}
\qqq
for all triangles $\,c$, \,all edges $\,b$, \,and for all vertices $\,v\,$
except for the TRIM for which such a choice would not be possible. Separating 
the contributions to $\,\Hol_\CG(\widetilde w)\,$ on the right hand side 
of (\ref{Hol}) into the ones coming from $\,\mathbb T^2_+\,$ and  
$\,\mathbb T^2_-$, \,we obtain
\qq
\Hol_\CG(\widetilde w)\,=\bigg(\hspace{-0.1cm}\exp\Big[\ii\sum\limits_{c\subset\mathbb T^2_+}
\int_c\widetilde w^*B_{i_c}\Big]\hspace{-0.1cm}
\mathop{\otimes}\limits_{\substack{(b,c)\\b\subset c\subset\mathbb T^2_+}}
\hspace{-0.2cm}\hol_{L_{i_ci_b}}(\widetilde w|_b)\bigg)\otimes\bigg(
\hspace{-0.1cm}\exp\Big[\ii\sum\limits_{c\subset\mathbb T^2_-}
\int_c\widetilde w^*B_{i_c}\Big]\hspace{-0.1cm}
\mathop{\otimes}\limits_{\substack{(b,c)\\b\subset c\subset\mathbb T^2_-}}
\hspace{-0.2cm}\hol_{L_{i_ci_b}}(\widetilde w|_b)\bigg).\qquad
\label{Hol1}
\qqq
\begin{figure}[t]
\begin{center}
\leavevmode
\hspace{0.9cm}
 \includegraphics[width=5.8cm,height=4.8cm]{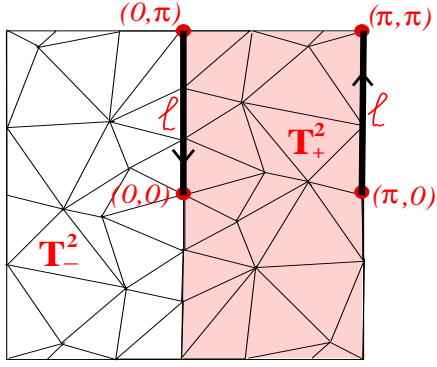}
\vskip -0.1cm
\caption{Triangulation of $\,\mathbb T^2\,$ 
for the calculation of $\,\Hol_\CG(\widetilde w)$}
\label{fig:triang}
\end{center}
\vskip -0.2cm
\end{figure} 
\vskip -0.2cm
\noindent There are several cancellations between the two contributions
on the right hand side. First, due to (\ref{rwt}), (\ref{iii}) and (\ref{BrB}),
\qq
&&\exp\Big[\ii\sum\limits_{c\subset\mathbb T^2_-}
\int_c\widetilde w^*B_{i_c}\Big]=
\exp\Big[\ii\sum\limits_{c\subset\mathbb T^2_+}
\int_{\vartheta(c)}\widetilde w^*B_{i^r_c}\Big]
=\exp\Big[\ii\sum\limits_{c\subset\mathbb T^2_+}
\int_c\vartheta^*\widetilde w^*B_{i^r_c}\Big]\cr
&&\hspace{1.5cm}=\exp\Big[\ii\sum\limits_{c\subset\mathbb T^2_+}
\int_c\widetilde w^*r^*B_{i^r_c}\Big]
=\exp\Big[-\ii\sum\limits_{c\subset\mathbb T^2_+}
\int_c\widetilde w^*B_{i_c}\Big]
\qqq
so that the integrals over the triangles cancel out
reducing formula (\ref{Hol1}) to
\qq
\Hol_\CG(\widetilde w)\,=\bigg(
\mathop{\otimes}\limits_{\substack{(b,c)\\b\subset c\subset\mathbb T^2_+}}
\hspace{-0.2cm}\hol_{L_{i_ci_b}}
(\widetilde w|_b)\bigg)\otimes\bigg(
\mathop{\otimes}\limits_{\substack{(b,c)\\b\subset c\subset\mathbb T^2_-}}
\hspace{-0.2cm}\hol_{L_{i_ci_b}}
(\widetilde w|_b)\bigg).
\label{Hol2}\qqq
Now note that, \,as explained in Sec\,\ref{sec:gerbeholonomy},  
\qq
\mathop{\otimes}\limits_{\substack{(b,c)\\b\subset c
\subset\mathbb T^2_+}}\hspace{-0.2cm}\hol_{L_{i_ci_b}}(\widetilde w|_b)\,\in\,
(\CL^\CG)_{\widetilde w|_{\partial\mathbb T^2_+}},
\qquad\mathop{\otimes}\limits_{\substack{(b,c)\\b\subset c
\subset\mathbb T^2_-}}\hspace{-0.2cm}\hol_{L_{i_ci_b}}(\widetilde w|_b)\,\in\,
(\CL^\CG)_{\widetilde w|_{\partial\mathbb T^2_-}}
=\,(\CL^\CG)^{-1}_{\widetilde w|_{\partial\mathbb T^2_+}},
\label{lrr}
\qqq
see (\ref{rhtm}), \,permitting to view the right hand side 
of (\ref{Hol2}) as a number in a way consistent with the  
previous such interpretation based on the subsequent use 
of maps $\,t_{ijk}$. From (\ref{rwt}), (\ref{iii}) and (\ref{holnu})
we further obtain:
\qq
&&\hspace{-1.3cm}(\CL^\CG)^{-1}_{\widetilde w|_{\partial\mathbb T^2_+}}\,\ni\,
\mathop{\otimes}\limits_{\substack{(b,c)\\b\subset c\subset\mathbb T^2_-}}
\hspace{-0.2cm}\hol_{L_{i_ci_b}}(\widetilde w|_b)\,
=\hspace{-0.1cm}\mathop{\otimes}\limits_{\substack{(b,c)\\b\subset 
c\subset\mathbb T^2_+}}\hspace{-0.2cm}
\hol_{L_{i^r_ci^r_b}}(\widetilde w|_{\vartheta(b)})\cr
&&\hspace{0.4cm}
=\mathop{\otimes}\limits_{\substack{(b,c)\\b\subset c\subset\mathbb T^2_+}}
\hspace{-0.2cm}\hol_{L_{i^r_ci^r_b}}
(\widetilde w\circ\vartheta|_b)\,
=\mathop{\otimes}\limits_{\substack{(b,c)\\b\subset c\subset\mathbb T^2_+}}
\hspace{-0.2cm}\hol_{L_{i^r_ci^r_b}}(r\circ\widetilde w|_b)
\,=\,\nu\Big(\hspace{-0.1cm}\mathop{\otimes}\limits_{\substack{(b,c)\\b\subset c
\subset\mathbb T^2_+}}\hspace{-0.2cm}\hol_{L_{i_bi_c}}(\widetilde w|_b)\Big).
\label{llr}
\qqq
Let us represent $\,\partial\mathbb T^2_+\,$ as $\,\ell\cup\vartheta(\ell)$,
\,where $\,\ell\,$ is the union of two closed vertical intervals
between the TRIM $\,(\pi,0)$,\ $\,(\pi,\pi)\,$ and $\,(0,\pi)$,\ $(0,0)$,
\,respectively, \,see Fig.\,1. For $\,v\in\ell$, \,we shall write
\qq
\widetilde w(v)=\gamma_v\ee^{2\pi\ii\tau_v}\gamma_v^{-1}\in\mathop{\cap}
\limits_{\substack{b\subset\ell\\v\in b}}O_{i_vi_b}
\qqq
so that $\,\widetilde w(\vartheta(v))=r(\widetilde w(v))
=(\gamma_v^{-1})^T\omega^{-1}\ee^{2\pi\ii\tau_v^r}\omega\gamma_v^T$,
\,see (\ref{r(g)}). By the left of relations (\ref{lrr}), the definition
of the line $\,(\CL^\CG)_{\widetilde w|_{\partial\mathbb T^2_+}}$, \,see (\ref{defline}) 
and (\ref{rhtm}), and the $\,\vartheta$-symmetry of the triangulation 
of $\,\partial T_+\,$ together with (\ref{iii}), 
\qq
\mathop{\otimes}\limits_{\substack{(v,b)\\v\in b\subset\partial\mathbb T^2_+}}
\hspace{-0.1cm}(L_{i_vi_b})_{\widetilde w(v)}^{\pm1}\,\ni\,
\mathop{\otimes}\limits_{\substack{(b,c)\\b\subset c
\subset\mathbb T^2_+}}\hspace{-0.2cm}\hol_{L_{i_ci_b}}(\widetilde w|_b)\ &=&
\mathop{\otimes}\limits_{\substack{(v,b)\\v\in b\subset\ell\\v\not\in\partial\ell}}
\Big(\big([\gamma_v,\zeta_v]_{i_vi_b}^{\pm1}\otimes
[(\gamma_v^{-1})^T\omega^{-1},\zeta'_v]_{i_v^ri_b^r}^{\mp1}
\big)\Big)\cr
&\otimes&\Big(\mathop{\otimes}\limits_{\substack{(v,b)\\v\in b\subset\ell\\v\in\partial\ell}}\big([\gamma_v,\zeta_v]_{i_vi_b}]^{\pm1}\otimes
[(\gamma_v^{-1})^T\omega^{-1},\zeta'_v]_{i_vi_b^r}^{\mp1}
\big)\Big)
\label{f1}
\qqq
for some $\,\zeta_v,\zeta'_v\in\mathbb C$, \,where $\,[\gamma,\zeta]_{ij}^{-1}\,$
denotes the element of $\,L_{ij}^{-1}\,$ dual to $\,[\gamma,\zeta]_{ij}\,$ in
$\,L_{ij}\,$ that may be identified with $\,[\gamma,\zeta^{-1}]_{ji}\in L_{ji}$.  
\,Similarly, \,from (\ref{llr}),
\qq
\mathop{\otimes}\limits_{\substack{(v,b)\\v\in b\subset\partial\mathbb T^2_+}}
\hspace{-0.2cm}(L_{i_vi_b})_{\widetilde w(v)}^{\mp1}\,\ni\hspace{-0.1cm}
\mathop{\otimes}\limits_{\substack{(b,c)\\b\subset c
\subset\mathbb T^2_-}}\hspace{-0.2cm}\hol_{L_{i_ci_b}}(\widetilde w|_b)\ &=&
\nu\bigg(\mathop{\otimes}\limits_{\substack{(v,b)\\v\in b\subset\ell\\
v\not\in\partial\ell}}
\Big(\big([\gamma_v,\zeta_v^{-1}]_{i_bi_v}^{\pm1}\otimes
[(\gamma_v^{-1})^T\omega^{-1},\zeta'^{-1}_v]_{i_b^ri_v^r}^{\mp1}
\big)\Big)\cr
&&\quad\ \otimes\,\Big(\mathop{\otimes}\limits_{\substack{(v,b)\\
v\in b\subset\ell\\v\in\partial\ell}}\big([\gamma_v,\zeta_v^{-1}]_{i_bi_v}]^{\pm1}
\otimes[(\gamma_v^{-1})^T\omega^{-1},\zeta'^{-1}_v]_{i_b^ri_v}^{\mp1}
\big)\Big)\bigg).\quad\qquad
\qqq
Upon using (\ref{nu}) and the relation
\qq
(((\gamma^{-1})^T\omega^{-1})^{-1})^T\omega^{-1}=(\omega\gamma^T)^T\omega^{-1}=
\gamma\,\omega^T\omega^{-1}=\gamma\,(-I)\,,
\qqq  
this gives
\qq
\mathop{\otimes}\limits_{\substack{(b,c)\\b\subset c
\subset\mathbb T^2_-}}\hspace{-0.2cm}\hol_{L_{i_ci_b}}(\widetilde w|_b)\ &=&
\Big(\mathop{\otimes}\limits_{\substack{(v,b)\\v\in b\subset\ell\\
v\not\in\partial\ell}}\big([(\gamma_v)^{-1})^T\omega^{-1},
\zeta_v^{-1}]_{i_v^ri_b^r}^{\pm1}\otimes
[\gamma_v(-I),\zeta'^{-1}_v]_{i_vi_b}^{\mp1}
\big)\Big)\cr
&\otimes&\Big(\mathop{\otimes}\limits_{\substack{(v,b)\\v\in b\subset\ell\\
v\in\partial\ell}}\big([(\gamma_v^{-1})^T\omega^{-1},\zeta_v^{-1}]_{i_v^ri_b^r}]
^{\pm1}\otimes
[\gamma_v(-I),\zeta'^{-1}_v]_{i_v^ri_b}^{\mp1}
\big)\Big).\qquad
\qqq
Since $\,\chi_{ij}(-I)=(-1)^{i-j}$, \,as may be easily seen from (\ref{tdm}), 
this can be rewritten using (\ref{eqrel}) as
\qq
\mathop{\otimes}\limits_{\substack{(b,c)\\b\subset c
\subset\mathbb T^2_-}}\hspace{-0.2cm}\hol_{L_{i_ci_b}}(\widetilde w|_b)\ &=&
\Big(\mathop{\otimes}\limits_{\substack{(v,b)\\v\in b\subset\ell\\
v\not\in\partial\ell}}\big([(\gamma_v)^{-1})^T\omega^{-1},
\zeta_v^{-1}]_{i_v^ri_b^r}^{\pm1}\otimes
[\gamma_v,(-1)^{i_v-i_{b}}\zeta'^{-1}_v]_{i_vi_b}^{\mp1}
\big)\Big)\cr
&\otimes&\Big(\mathop{\otimes}\limits_{\substack{(v,b)\\v\in b\subset\ell\\
v\in\partial\ell}}\big([(\gamma_v^{-1})^T\omega^{-1},\zeta_v^{-1}]_{i_v^ri_b^r}]
^{\pm1}\otimes
[\gamma_v,(-1)^{i_v+m-i_b}\zeta'^{-1}_v]_{i_v^ri_b}^{\mp1}
\big)\Big)\cr\cr\cr
&=&
\Big(\mathop{\otimes}\limits_{\substack{(v,b)\\v\in b\subset\ell\\
v\not\in\partial\ell}}\big([(\gamma_v)^{-1})^T\omega^{-1},
\zeta_v^{-1}]_{i_v^ri_b^r}^{\pm1}\otimes
[\gamma_v,\zeta'^{-1}_v]_{i_vi_b}^{\mp1}
\big)\Big)\cr
&\otimes&\Big(\mathop{\otimes}\limits_{\substack{(v,b)\\v\in b\subset\ell\\
v\in\partial\ell}}\big([(\gamma_v^{-1})^T\omega^{-1},\zeta_v^{-1}]_{i_v^ri_b^r}]
^{\pm1}\otimes
[\gamma_v,(-1)^{i_v}\zeta'^{-1}_v]_{i_v^ri_b}^{\mp1}
\big)\Big).
\label{f2}
\qqq
The factors $\,(-1)^{i_b}\,$ canceled because each $\,b\subset\ell\,$ appeared
two times, and similarly for the factors $\,(-1)^{i_v}\,$ for $\,v\in\ell$, 
$\,v\not\in\partial\ell$. The factors $\,(-1)^m\,$ also disappeared 
as they occurred four times. \,Substituting (\ref{f1}) and (\ref{f2}) to 
(\ref{Hol2}) and using the duality between pairs of factors in the tensor 
product, we observe that all the contributions from $\,v\not\in\partial\ell\,$ 
cancel out so that
\qq
\Hol_\CG(\widetilde w)\,&=&\bigg(\mathop{\otimes}
\limits_{\substack{(v,b)\\v\in b\subset\ell\\v\in\partial\ell}}\big([\gamma_v,\zeta_v]_{i_vi_b}]^{\pm1}\otimes
[(\gamma_v^{-1})^T\omega^{-1},\zeta'_v]_{i_vi_b^r}^{\mp1}
\big)\bigg)\cr
&&\hspace{-0.26cm}\otimes\bigg(
\mathop{\otimes}\limits_{\substack{(v,b)\\v\in b\subset\ell\\
v\in\partial\ell}}\big([(\gamma_v^{-1})^T\omega^{-1},\zeta_v^{-1}]_{i_v^ri_b^r}]
^{\pm1}\otimes
[\gamma_v,(-1)^{i_v}\zeta'^{-1}_v]_{i_v^ri_b}^{\mp1}
\big)\bigg)\cr\cr\cr
&=&\mathop{\otimes}\limits_{\substack{(v,b)\\v\in b\subset\ell\\v\in\partial\ell}}\big([\gamma_v,(-1)^{i_v}\zeta_v\zeta'_v]_{i_vi_v^r}]^{\pm1}\otimes
[(\gamma_v^{-1})^T\omega^{-1},\zeta_v\zeta'_v]_{i_vi_v^r}^{\mp1}
\big),
\qqq
where the last line was obtained using the isomorphisms $\,t_{i_vi_bi_v^r}\,$
and $\,t_{i_vi_b^ri_v^r}$. \,But $\,v\in\partial\ell\,$ are TRIM 
for which $\,\widetilde w(v)=r(\widetilde w(v))\,$ implying that
\qq
\gamma_v\ee^{2\pi\ii\tau_v}\gamma_v=(\gamma_v^{-1})^T\omega^{-1}\ee^{2\pi\ii\tau_v^r}
\omega\gamma_v^T\,\in\,O_{i_vi_v^r}
\qqq
and, consequently, that $\,\tau=\tau^r\,$ and  
\qq
\gamma_{v0}=\gamma_v^{-1}(\gamma_v^{-1})^T\omega^{-1}\,\in\,G_{i_vi_v^r}\,.
\qqq
Hence, \,by (\ref{eqrel}),
\qq
\Hol_\CG(\widetilde w)\,&=&
\mathop{\otimes}\limits_{\substack{(v,b)\\v\in b\subset\ell\\v\in\partial\ell}}
\big([\gamma_v,(-1)^{i_v}\zeta_v\zeta'_v]_{i_vi_v^r}]^{\pm1}\otimes
[\gamma_v\gamma_{v0},\zeta_v\zeta'_v]_{i_vi_v^r}^{\mp1}\big)\cr\cr
&=&\mathop{\otimes}\limits_{\substack{(v,b)\\v\in b\subset\ell\\v\in\partial\ell}}
\big([\gamma_v,(-1)^{i_v}\zeta_v\zeta'_v]_{i_vi_v^r}]^{\pm1}\otimes
[\gamma_v,\chi_{i_vi_v^r}(\gamma_{v0})\zeta_v\zeta'_v]_{i_vi_v^r}^{\mp1}\big)\cr\cr
&=&\mathop{\otimes}\limits_{\substack{(v,b)\\v\in b\subset\ell\\v\in\partial\ell}}
(-1)^{i_v}\chi_{i_vi_v^r}(\gamma_{v0})^{\mp1}\,.
\label{f3}
\qqq
\vskip 0.2cm

\noindent{\bf Lemma 2.} \ For an antisymmetric $SU(n)$ matrix 
$\,\widetilde w=\gamma\ee^{2\pi\ii\tau}\gamma^{-1}=
r(\widetilde w)=(\gamma^{-1})^T\omega^{-1}\ee^{2\pi\ii\tau}\omega\gamma^T\in 
O_{ii^r}\,$ and for $\,\gamma_0=\gamma^{-1}(\gamma^{-1})^T\omega^{-1}$,
\qq
\pf\,{\widetilde w}\,=\,\ii^{m^2}(-1)^{i}\chi_{ii^r}(\gamma_0)\,.
\label{lemma2}
\qqq
\vskip 0.4cm

\noindent Assuming Lemma 2, the substitution of (\ref{lemma2}) for 
$\,\widetilde w=\widetilde w(v)\,$ and $\,v\in\partial\ell\,$ to the right 
hand side of (\ref{f3}) permits to complete the proof of Proposition 1 
\,(recall that $\,\pf\,{\widetilde w(v)}=\pm1$). 
\vskip 0.4cm

\noindent{\bf Proof of Lemma 2.} \ Let us first suppose that 
$\,\widetilde w\in O_{0m}$. \,Since
$\,\widetilde w=\gamma\,\ee^{2\pi\ii\tau}\gamma_0\omega\gamma^T\,$ and
$\,\gamma\in SU(n)$, \,it follows
that $\,\ee^{2\pi\ii\tau}\gamma_0\omega\,$ is an antisymmetric matrix and that
\qq
\pf\,{\widetilde w}=\det(\gamma)\,\pf(\ee^{2\pi\ii\tau}\gamma_0\omega)=
\pf(\ee^{2\pi\ii\tau}\gamma_0\omega)\,.
\label{cosc}
\qqq
Under the assumption that $\,\widetilde w\in O_{0m}$, 
$\,\tau=\sum\limits_{j=0}^{n-1}\tau_j\lambda_j\,$ with 
$\,\tau_{j^r}=\tau_j\,$ and $\,\tau_0=\tau_m>0$. \,Due to (\ref{ccos}),
\qq
\omega\tau\omega^{-1}=\sum\limits_{j=0}^{n-1}\tau_j\lambda_{j^r}-
\lambda_{m}
=\tau-\lambda_{m}\,.
\label{oto}
\qqq
Recall that $\,\lambda_m={\rm diag}[\frac{1}{2},\dots,\frac{1}{2},-\frac{1}{2},\dots,
-\frac{1}{2}]$.
\,If $\,2\pi\tau={\rm diag}[\varphi_1,\dots,\varphi_{n}]\,$ then (\ref{oto})
means that
\qq
{\rm diag}[\varphi_{m+1},\dots,\varphi_{n},\varphi_1,
\dots,\varphi_{m}]\,=\,
{\rm diag}[\varphi_1-\pi,\dots,\varphi_{m}-\pi,
\varphi_{m+1}+\pi,\dots,\varphi_{n}
+\pi]\,,
\qqq
i.e. $\,(\varphi_{m+1},\dots,\varphi_{n})
=(\varphi_1-\pi,\dots\varphi_{m}-\pi)$.
It follows that 
\qq
\sum\limits_{j=1}^{m}\varphi_j\,=\,\frac{_{\pi m}}{^2}\,.
\qqq
Besides, as $\,\gamma_0\in G_{0m}\,$ and $\,\lambda_{0m}
=\lambda_{m}$, \,the matrix $\,\gamma_0\,$ has a block structure with 
$m\times m$ blocks:
\qq
\gamma_0\,=\bigg(\begin{matrix}\gamma'_0&0\cr 0&\gamma^{\prime\prime}_0
\end{matrix}\bigg)
\qqq
and $\,\chi_{0m}(\gamma_0)=\det{\gamma'_0}=\det{\gamma^{\prime\prime}_0}^{-1}\,$
as is easy to see from (\ref{chiij}).
\,Since 
\qq
\gamma^{-1}(\gamma^{-1})^T=\gamma_0\omega
=\,\bigg(\begin{matrix}0&\gamma'_0\cr -\gamma^{\prime\prime}_0&0
\end{matrix}\bigg)
\qqq
is a symmetric matrix, \,it follows that $\,\gamma^{\prime\prime}_0=-(\gamma'_0)^T$.
Hence
\qq
\ee^{2\pi\ii\tau}\gamma_0\omega&=&\bigg(\begin{matrix}0&
{\rm diag}[\ee^{i\varphi_1},\dots,\ee^{\ii\varphi_{m}}]\gamma'_0\cr
{\rm diag}[\ee^{i\varphi_{m+1}},\dots,\ee^{\ii\varphi_{n}}]
(\gamma'_0)^T&0\end{matrix}\bigg)\cr\cr
&=&\bigg(\begin{matrix}0&
{\rm diag}[\ee^{i\varphi_1},\dots,\ee^{\ii\varphi_{m}}]\gamma'_0\cr
-{\rm diag}[\ee^{i\varphi_1},\dots,\ee^{\ii\varphi_{m}}]
(\gamma'_0)^T&0\end{matrix}\bigg)\cr\cr
&=&\bigg(\begin{matrix}0&
{\rm diag}[\ee^{i\varphi_1},\dots,\ee^{\ii\varphi_{m}}]\gamma'_0\cr
-({\rm diag}[\ee^{i\varphi_1},\dots,\ee^{\ii\varphi_{m}}]\gamma'_0)^T
&0\end{matrix}\bigg)
\qqq
and from the standard formula for the pfaffian of a block off-diagonal 
antisymmetric matrix,
\qq
\pf(\ee^{2\pi\ii\tau}\gamma_0\omega)&=&(-1)^{m(m-1)/2}
\det({\rm diag}[\ee^{i\varphi_1},
\dots,\ee^{\ii\varphi_{m}}]\gamma'_0)\,=\,(-1)^{m(m-1)/2}
\ee^{\ii(\varphi_1+\cdots+\varphi_{m})}\det(\gamma_0')\cr\cr
&=&(-1)^{m(m-1)/2}\ii^{m}\det(\gamma_0')\,=\,
\ii^{m^2}\det(\gamma'_0)\,=\,
\ii^{m^2}\chi_{0m}(\gamma_0)\,,
\qqq
which, together with (\ref{cosc}), \,proves (\ref{lemma2}) for $\,i=0$.
\vskip 0.1cm

Suppose now that $\,\widetilde w\in O_{ii^r}$. \,In this case,
$\,\widetilde w'=z^{-i}\widetilde w=-(\widetilde w')^T\in O_{0m}\,$
for $\,z=\ee^{-2\pi\ii/n}\,$ being the generator of the center of $\,SU(n)$.
\,Let $\,\omega_1\,$ be the $n\times n$ matrix in $\,SU(n)\,$ such that 
$\,\omega_1 e_l=\ee^{\pi\ii/n}e_{l+1}\,$ in the action on the vectors 
of the canonical basis of $\,\mathbb C^{n}$ (with the identification 
$\,e_{n+1}=e_1$). \,Then
\qq
&&\omega_1\lambda_j\omega_1^{-1}=\lambda_{j+1}-\lambda_1\,,\qquad
\omega_1^i\lambda_j\omega_1^{-i}=\lambda_{j+i}-\lambda_i\,,\qquad
\omega=\ee^{-\pi\ii\lambda_{m}}\omega_1^{m}\,,\\\cr
&&\omega_1^i\,\ee^{2\pi\ii\sum\limits_{j=0}^{n-1}\tau_j\lambda_{j-i}}\omega_1^{-i}
=\ee^{2\pi\ii\sum\limits_{j=0}^{n-1}\tau_j\lambda_j-2\pi\ii\lambda_i}=z^{-i}
\,\ee^{2\pi\ii\tau}\,,
\qqq
where $\,j\,$ in $\,\lambda_j\,$ is taken modulo $\,n$. 
Hence, if $\,\widetilde w=\gamma\,\ee^{2\pi\ii\tau}\gamma^{-1}\,$ then
$\,\widetilde w'=\gamma\,\omega_1^i\,\ee^{2\pi\ii\sum\limits_j\tau_j\lambda_{j-i}}
\omega^{-i}\gamma^{-1}$. \,Let
\qq
\gamma'_0&=&(\gamma\omega_1^i)^{-1}((\gamma\omega_1^i)^{-1})^T\omega^{-1}\,=\,
\omega_1^{-i}\gamma^{-1}(\gamma^{-1})^T((\omega_1^{-1})^T)^i\omega_1^{-m}
\ee^{\pi\ii\lambda_{m}}\cr\cr
&=&\ee^{-\pi\ii\, i/m}\,\omega_1^{-i}\gamma^{-1}(\gamma^{-1})^T
\omega_1^i\omega_1^{-m}\ee^{\pi\ii\lambda_{m}}
\,=\,\ee^{-\pi\ii\, i/m}\,\omega_1^{-i}\gamma^{-1}(\gamma^{-1})^T\omega^{-1}
\ee^{-\pi\ii\lambda_{m}}\omega_1^i\ee^{\pi\ii\lambda_{m}}\cr\cr
&=&\ee^{-\pi\ii\, i/m}\,\omega_1^{-i}\gamma_0
\,\ee^{-\pi\ii\lambda_{m}}\omega_1^i\ee^{\pi\ii\lambda_{m}}\,.
\qqq
In other words, $\,\gamma'_0\in G_{0m}\,$ satisfies the relation
\qq
\ee^{\pi\ii\, i/m}\,\gamma_0'\,\ee^{-\pi\ii\lambda_{m}}\,=\,\omega_1^{-i}\gamma_0
\,\ee^{-\pi\ii\lambda_{m}}\omega_1^i\,.
\label{tbcc}
\qqq
As the adjoint action of $\,\omega_1^i\,$ sends $\,\lambda_{0m}$ to 
$\,\lambda_{ii^r}$, \,it maps $\,G_{0m}\,$ into $\,G_{ii^r}\,$ and 
intertwines the characters $\,\chi_{0m}\,$ and $\,\chi_{ii^r}$.
\,From (\ref{tbcc}), \,it follows then that
\qq
\chi_{0m}(\ee^{\pi\ii\, i/m}\,\gamma_0'\,\ee^{-\pi\ii\lambda_{m}})\,=\,
(-1)^i\chi_{0m}(\gamma_0'\,\ee^{-\pi\ii\lambda_{m}})\,=\,
\chi_{ii^r}(\gamma_0\,\ee^{-\pi\ii\lambda_{m}})\,.
\label{ala}
\qqq
Relation (\ref{tdm}) and the equalities
$\,\tr(\lambda_i\lambda_{m})
=\Big\{\substack{i/2\\ (n-i)/2}\Big\}\,$ for 
$\,\Big\{\substack{\,i\leq m\\ \,i\geq m}\Big\}\,$
imply that
\qq
\chi_{0m}(\ee^{-\pi\ii\lambda_{m}})\,=\,\ee^{-\pi\ii m/2}\,=\,\ii^{-m}\,,\qquad
\chi_{ii^r}(\ee^{-\pi\ii\lambda_{m}})\,=\,\ee^{\pi\ii\,\tr((\lambda_i-\lambda_{i^r})
\lambda_{m})}\,=\,(-1)^i\ii^{-m}\,.
\qqq
Hence (\ref{ala}) reduces to the identity
\qq
\chi_{0m}(\gamma_0')\,=\,\chi_{ii^r}(\gamma_0)\,.
\qqq
Using (\ref{lemma2}) for $\,i=0$, \,we infer then that
\qq
\ii^{-m^2}\,\pf(\widetilde w')\,=\,\chi_{0m}(\gamma_0')\,=\,
\chi_{ii^r}(\gamma_0)\,.
\qqq
But $\,\pf\,{\widetilde w'}=\pf(\ee^{2\pi\ii\, i/n}\widetilde w)
=(-1)^i\,\pf\,{\widetilde w}\,$ and (\ref{lemma2}) for general 
$\,i\,$ follows completing the proof of Lemma 2.

\nsection{$3d$ Fu-Kane-Mele invariant and the Chern Simons action of the Berry
connection}
\label{sec:FKMandCS}

\noindent We shall show here that the relation between the strong 
Fu-Kane-Mele invariant $\,\KM^s\in\mathbb Z_2\,$ for $\,3d\,$ TRI topological 
insulators \cite{FKM} and the Chern-Simons (CS) action of the Berry connection 
of the valence bundle, first rigorously established in \cite{FM},
\,is a corollary of the Theorem obtained in the previous sections. 
The invariant $\,\KM^s\,$ was defined in \cite{FKM} by the formula 
\qq
(-1)^{\KM^s}=\prod\limits_{{\rm TRIM}\,\in\,\mathbb T^3}
\frac{\sqrt{\det{w(k)}}}{\pf\,{w(k)}}
\label{FKM3}
\qqq 
similar to (\ref{FKM2}) but with the product over the $8$ TRIM in the 
$3d$ Brillouin torus. The non-Abelian Berry connection $\,A\,$ of the 
valence bundle $\,\CE\,$ over $\,\mathbb T^3\,$ is defined using an 
orthonormal frame $\,(e_i(k)),\ i=1,\dots,n$, \,globally trivializing 
$\,\CE$. It is a matrix-valued 1-form on $\,\mathbb T^3\,$ given by 
the formula
\qq
A_{ij}(k)\,=\,\big\langle e_i(k)|de_j(k)\big\rangle=-\overline{A_{ji}(k)}\,.
\qqq
It determines the CS 3-form
\qq
\SC(A)\,=\,\frac{_1}{^{4\pi}}\,\tr\big(AdA+\frac{_2}{^3}A^3\big)
\qqq
and the CS action 
\qq
S_\SC(A)\,=\,\int_{\mathbb T^3}\SC(A)\,.
\label{CSaction}
\qqq
Under the change of the trivializing frame,
\qq
e_{i}(k)=\sum\limits_iU(k)_{i'i}\,e'_i(k)\,,
\qqq
the connection form transforms by the gauge transformation 
$\,A'=UAU^{-1}+UdU^{-1}\,$ and the CS 3-form by
\qq
\SC(A')=\SC(A)+U^*H+\frac{_1}{^{4\pi}}d(\tr(U^{-1}dU)A\,,
\label{TrCS}
\qqq
where the 3-form $\,H\,$ on $\,U(n)\,$ is given by (\ref{H}).
\,The transformation rule (\ref{TrCS}) results in the relation
\qq
S_\SC(A')\,=\,S_\SC(A)\,+\,\int_{\mathbb T^3}U^*H
\qqq
which implies that the CS action $\,S_\SC(A)\,$ changes by multiples of
$2\pi$ under gauge transformations (recall that the $3$-periods of $H$ are
in $\,2\pi\mathbb Z$) \,and that the CS Feynman amplitude 
$\,\exp[\ii S_\SC(A)]\,$ is gauge invariant. We shall show the following 
result:
\vskip 0.5cm

\noindent{\bf Proposition 2.}
\vskip -1.1cm

\qq
(-1)^{\KM^s}\,=\,\exp\Big[\ii S_\SC(A)\Big].
\label{FKMCS}
\qqq
\vskip 0.3cm

\noindent{\bf Proof.} \ The argument below closely follows \cite{QHZ}, using 
at the end our Theorem from Sec.\,\ref{sec:mainresult}. Let us start by taking 
$\,e'_{i'}(k)=\theta e_i(-k)\,$ as the new trivialization of $\,\CE$. 
\,One has
\qq
e_i(k)&=&\sum\limits_{i'}\big\langle \theta e_{i'}(-k)|e_i(k)\big\rangle\,
\theta e_{i'}(-k)\,=\,\sum\limits_{i'}\overline{\big\langle e_i(k)|
\theta e_{i'}(-k)\big\rangle}\,e'_{i'}(k)\cr&=&\sum\limits_{i'}
\overline{w_{ii'}(-k)}\,e'_{i'}(k)\,=\,-\sum\limits_{i'}
\overline{w_{i'i}(k)}\,e'_{i'}(k)\,,
\qqq 
where we used the symmetry (\ref{TRw}) of $\,w$. It follows from the previous 
discussion that $\,A'\,$ is a gauge transformation of $\,A$,
\qq
A'=\overline{w}A\overline{w}^{-1}+\overline{w}d\overline{w}^{-1}\,,
\qqq
and, from (\ref{TrCS}), that
\qq
\SC(A')\,=\,\SC(A)+\overline{w}^*H
+\frac{_1}{^{4\pi}}\,d\,\tr\big((\overline{w}^{-1}d\overline{w})A\big)
\,=\,\SC(A)+w^*H
+\frac{_1}{^{4\pi}}\,d\,\tr(A^Tw^{-1}dw)\,.
\label{CS'}
\qqq
But, by the antiunitarity of $\,\theta$,
\qq
A'_{i'j'}(k)\,=\,\langle\theta e_{i'}(-k)|d\theta e_{j'}(-k)\rangle\,=\,
\langle de_{j'}(-k)|e_{i'}(-k)\rangle\,
=\,-\langle e_{j'}(-k)|de_{i'}(-k)\rangle\,=\,-A_{j'i'}(-k)\,.
\qqq
or $\,A'=-\vartheta^*A^T\,$ for $\,\vartheta:\mathbb T^3\rightarrow
\mathbb T^3\,$ induced by $\,k\mapsto -k$. \,This implies that 
$\,\SC(A')=\vartheta^*\SC(A)\,$ and, with the use
of (\ref{CS'}), that
\qq
\vartheta^*\SC(A)\,=\,\SC(A)+w^*H
+\frac{_1}{^{4\pi}}\,d\,\tr(A^Tw^{-1}dw)\,.
\qqq
Integrating the latter identity over $\,\mathbb T^3$, \,remembering that $\,\vartheta\,$ reverses 
the orientation of $\,\mathbb T^3$, \,we obtain the relation \cite{QHZ}
\qq
S_\SC(A)\,=\,-\frac{_1}{^2}\int_{\mathbb T^3}w^*H\,.
\qqq
Let, similarly as in $2d$, $\,\mathbb T^3_{\pm}\,$ correspond to the parts of
$\,\mathbb T^3\,$ with $\,\pm k_1\in[0,\pi]$, \,see Fig\,2. \,Then
\qq
\int_{\mathbb T^3}w^*H=\int_{\mathbb T^3_+}w^*H+\int_{\mathbb T^3_-}w^*H=
\int_{\mathbb T^3_+}(w^*H-\vartheta^*w^*H)
=\int_{\mathbb T^3_+}(w^*H-(w^T)^*H)=2\int_{\mathbb T^3_+}w^*H\qquad
\label{onwH}
\qqq
using the symmetry (\ref{TRw}) and the relation $\,(w^T)^*H=-w^*H$.
\,Hence
\qq
S_\SC(A)\,=\,-\int_{\mathbb T^3_+}w^*H\,. 
\qqq
Exponentiating the latter identity, one finally obtains
\qq
\exp\Big[\ii S_\SC(A)\Big]\,=\,
\exp\Big[-\ii\int_{\mathbb T^3_+}w^*H\Big]\,=\,
\frac{\exp\big[\ii S_\WZ(w|_{k_1=0})\big]}
{\exp\big[\ii S_\WZ(w|_{k_1=\pi})\big]}\,,
\label{CS0pi}
\qqq
where the second equality follows easily from the definition
(\ref{WZ}) of the WZ action. Denote by $\,\mathbb T^2_a\,$ 
the two-dimensional subtorus of $\,\mathbb T^3\,$ corresponding to $\,k_1=a\,$
for $\,a=0,\pi$, \,see Fig.\,2. By our main Theorem of 
Sec.\,\ref{sec:mainresult},
\qq
\exp\big[\ii S_\WZ(w|_{k_1=a})\big]\,=\,\prod\limits_{{\rm TRIM}\in\mathbb T^2_{a}}
\frac{\sqrt{\det{w(k)}}}{\pf\,{w(k)}}\,.
\label{a0pi}
\qqq
Identity (\ref{FKM3}) follows now from (\ref{CS0pi}) and (\ref{a0pi}), completing 
the proof of Proposition 2.  
\begin{figure}[t]
\begin{center}
\leavevmode
\hspace{0.3cm}
 \includegraphics[width=7.1cm,height=6cm]{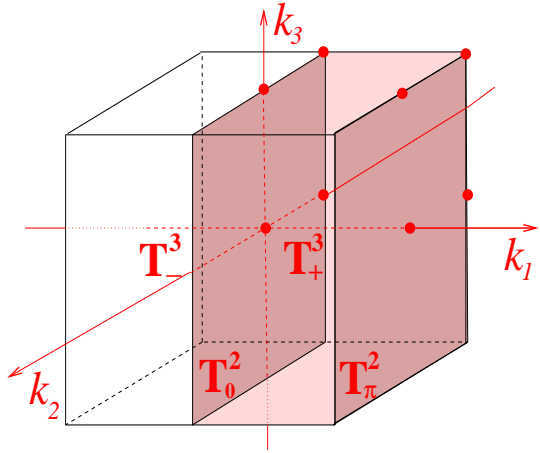}
\vskip -0.1cm
\caption{$3d\,$ Brillouin torus $\,\mathbb T^3$}
\label{fig:torus3d}
\end{center}
\vskip -0.2cm
\end{figure} 

\nsection{Conclusions}
\label{sec:concl}

\noindent We have proved that the Fu-Kane-Mele $\,\mathbb Z_2$-valued 
invariant of $2d$ time-reversal-symmetric crystalline insulators can 
be expressed as the properly normalized Wess-Zumino action of the sewing 
matrix field defined on the Brillouin torus $\,\mathbb T^2$. \,This was 
done by the localization of the WZ action in question at the four 
time-reversal-invariant (quasi-)momenta, obtained using the bundle-gerbe 
technique. Our result shed new light on the $2d$ Fu-Kane-Mele invariant 
and its topological nature. Applied in the $3d$ setup, it also permitted 
a direct proof of the relation between the strong Fu-Kane-Mele invariant 
of TRI crystals and the Chern-Simons action of the non-Abelian Berry 
connection on the bundle of valence states over the Brillouin torus 
$\,\mathbb T^3$. 
\vskip 0.1cm

Previously, the present author with collaborators established different 
relations between the $2d$ and $3d$ Fu-Kane-Mele invariants and 
the Wess-Zumino action \cite{CDFG,CDFGT,Geneva}. In particular, 
the $2d$ Fu-Kane-Mele invariant was described in terms of the properly 
defined square root of the Wess-Zumino amplitude of the unitary-group-valued 
field $\,U(k)=I-2P(k)\,$ and the strong $3d$ Fu-Kane-Mele invariant 
in terms of a related $\,\pm1$-valued index of the $3d$ version of the 
same field. Those constructions were extended to the case of periodically 
forced TRI crystalline systems allowing to define $\,\mathbb Z_2$-valued 
refinements of the $\,\mathbb Z$-valued dynamical indices that were
introduced in \cite{RLBL} for Floquet systems without TRI. As discussed 
in \cite{Geneva}, the geometric framework of those works involved 
$\mathbb Z_2$ equivariant structures on gerbes. A direct relation between 
the constructions presented there and the simpler ones described here is 
still missing. In particular, it is not clear whether it is possible 
to extend directly the static discussion of the present paper to the case 
of periodically forced crystals.

\nappendix{}
\renewcommand{\theequation}{A.\arabic{equation}}
\vskip 0.3cm

\noindent We establish here (in the reversed order) the properties 
of the WZ amplitude $\,\exp[\ii S_\WZ(w)]\,$ claimed in Remark 
in Sec.\,\ref{sec:mainresult}. 
\vskip 0.1cm

First, for a smooth family of 
fields $\,w_t(k)\,$ defined on $\,\mathbb T^2\,$ such that 
$\,w_t\circ\vartheta=-w_t^T\,$ for all $\,t\,$ (with $\,\vartheta\,$
induced by $\,k\mapsto-k$) \,the well known
formula for the derivative of the WZ action, easily following from 
the definition (\ref{WZ}), gives:
\qq
\frac{_d}{^{dt}}S_\WZ(w(t))&=&\frac{_1}{^{4\pi}}\int_{\mathbb T^2}\tr\big(
(w^{-1}_t\partial_tw_t)(w_t^{-1}dw_t)^2\big)
=\frac{_1}{^{4\pi}}\int_{\mathbb T^2}\vartheta^*
\tr\big((w^{-1}_t\partial_tw_t)(w_t^{-1}dw_t)^2\big)\cr
&=&\frac{_1}{^{4\pi}}\int_{\mathbb T^2}
\tr\big(((w^T_t)^{-1}\partial_tw_t^T)((w_t^T)^{-1}dw_t^T)^2\big)
=\frac{_1}{^{4\pi}}\int_{\mathbb T^2}
\tr\big(((w^T_t)^{-1}\partial_tw_t^T)((w_t^T)^{-1}dw_t^T)^2\big)^T\cr
&=&-\frac{_1}{^{4\pi}}\int_{\mathbb T^2}
\tr\big(((dw_t)w_t^{-1})^2(\partial_tw_t)w_t^{-1}\big)
=-\frac{_1}{^{4\pi}}\int_{\mathbb T^2}
\tr\big((w_t^{-1}\partial_tw_t)(w_t^{-1}dw_t)^2\big)\,=\,0\,,
\qqq
implying the invariance of the WZ amplitude $\,\exp[\ii S_\WZ(w)]\,$
under smooth deformations of $\,w\,$ preserving the symmetry (\ref{TRw}).
\vskip 0.1cm

Next, let us consider a change of the trivialization of the valence bundle
\qq
e_i(k)=\sum\limits_{i'}U_{i'i}(k)\,e'_{i'}(k)
\qqq
with unitary $\,U(k)$. \,For the sewing matrices, this gives: 
\qq
w_{ij}(k)=\big\langle e_i(-k)|\theta e_j(k)\big\rangle=
=\sum\limits_{i',j'}\overline{U_{i'i}(-k)\,U_{j'j}(k)}
\,\big\langle e'_{i'}(-k)|\theta e'_{j'}(k)\big\rangle=
\sum\limits_{i',j'}U^{-1}_{\ ii'}(-k)\,U^{-1}_{\ jj'}(k)\,w'_{i'j'}(k)\quad
\qqq
so that $\,w(k)=U^{-1}(-k)w'(k)(U^{-1}(k))^T\,$ or
\qq
w'(k)=U(-k)w(k)U(k)^T.
\qqq
The map $\,\mathbb T^2\ni k\mapsto U(k)\in U(n)\,$ can be 
smoothly contracted to the one
\qq
\mathbb T^2\ni k\mapsto U_{n_1,n_2}(k)={\rm diag}[\ee^{\ii(n_1k_1+n_2k_2)},1,
\dots,1]\,,
\qqq
where $\,n_1,n_2\in\mathbb Z\,$ are the winding numbers of $\,\det{U(k)}\,$
along the basic cycles of $\,\mathbb T^2$. \,By the previous argument,
it is enough to check the relation  $\,\exp[\ii S_\WZ(w')]
=\exp[\ii S_\WZ(w)]\,$ for $\,U(k)=U_{n_1,n_2}(k)$. \,Besides, by an
$\,SL(2,\mathbb Z)\,$ change of variables $\,k$, \,one may
achieve that $\,n_1=0$. \,Let $\,\CD\,$ be the unit disc in $\,\mathbb C$. 
\,Then $\,\CB=\CD\times S^1\,$ is a 3-manifold with the boundary 
$\,S^1\times S^1\cong\mathbb T^2$. \,Since $\,\det{w(k)}\,$ has
no windings, there exists a smooth contraction
\qq
[0,1]\times\mathbb T^2\ni(r,k)\,\longmapsto\, W(r,k)\in U(n)
\qqq
such that $\,W(1,k)=w(k)\,$ and $\,W(r,k)=I\,$ for $\,r\,$ close to zero. 
We shall identify $\,W\,$ with a smooth map defined on $\,\CB\,$ by setting
\qq
W(r\ee^{\ii k_1},\ee^{\ii k_2})=W(r,k)\,.
\qqq
Consider two other smooth maps $\,V_{1,2}:\CB\rightarrow U(n)\,$ given by
\qq
V_1(r\ee^{\ii k_1},\ee^{\ii k_2})={\rm diag}[\ee^{-\ii n_2k_2},1,\dots,1]\,,
\qquad V_2(r\ee^{\ii k_1},\ee^{\ii k_2})={\rm diag}[\ee^{\ii n_2k_2},1,\dots,1]\,.
\qqq
The product map $\,W'=V_1WV_2:\CB\rightarrow U(n)\,$ is a smooth extension 
of $\,w'(k)=U_{0,n_2}(-k)w(k)U_{0,n_2}(k)^T\,$ to the interior of 
$\,\CB\,$ so that, by Witten's prescription,
\qq
S_\WZ(w')&=&\int_\CB (W')^*H=\int_\CB(V_1WV_2)^*H\cr
&=&\int_\CB\Big(V_1^*H+
W^*H+V_2^*H+\frac{_1}{^{4\pi}}\,d\,\tr\big((V_1^{-1}dV_1)WV_2d(WV_2)^{-1}+
(W^{-1}dW)V_2dV_2^{-1}\big)\Big),\quad
\qqq
where we applied twice the formula
\qq
(W_1W_2)^*H=W_1^*H+W_2^*H+\frac{_1}{^{4\pi}}\,d\,\tr\big((W_1^{-1}dW_1)W_2dW_2^{-1}
\big)
\label{prfmla}
\qqq
holding for two $\,U(n)$-valued maps $\,W_{1,2}\,$ on the same
domain. \,Since $\,V_1^*H=0=V_2^*H\,$ for dimensional reasons, 
\,we infer that
\qq
&&S_\WZ(w')-S_\WZ(w)\cr
&&=\,\frac{_1}{^{4\pi}}\int_{\partial\CB}\tr\Big((V_1^{-1}dV_1)W(V_2dV_2^{-1})W^{-1}
-(V_1^{-1}dV_1)(dW)W^{-1}+(W^{-1}dW)V_2dV_2^{-1}\Big)\cr
&&=\,\frac{_1}{^{4\pi}}\int_{\mathbb T^2}\tr\Big({\rm diag}[-\ii n_2dk_2,0,\dots,0]
\,\,w\,\,{\rm diag}[-\ii n_2dk_2,0,\dots,0]\,w^{-1}\cr
&&\hspace{1.75cm}-{\rm diag}[-\ii n_2dk_2,0,\dots,0]\,(dw)w^{-1}
+(w^{-1}dw)\,{\rm diag}[-\ii n_2dk_2,0,\dots,0]\Big).
\qqq
Changing the variables $\,k\mapsto-k\,$ in the integral on the right 
hand side and using the symmetry (\ref{TRw}) of $\,w\,$ together with 
the invariance of trace under the transposition, \,one shows 
that the integral in question is equal to its negative, hence it vanishes.
\vskip 0.1cm

Finally, note that $\,\widetilde W(r,k)=-W(r,-k)^T\,$ is also a contraction of
$\,w(k)\,$ and, as $\,W$, \,it may be regarded as defined on $\,\CB$. \,Then
\qq
S_\WZ(w)=\int_\CB\tilde W^*H=\int_\CB(-W^T)^*H=-\int_\CB W^*H=-S_\WZ(w)
\qqq
modulo $\,2\pi$, \,implying \,that $\,\exp[\ii S_\WZ(w)]
=(\exp[\ii S_\WZ(w)])^{-1}=\pm1$.

\end{document}